\documentclass[conference]{IEEEtran}
\IEEEoverridecommandlockouts
\usepackage{cite}
\usepackage{amsmath,amssymb,amsfonts}
\usepackage{algorithmic}
\usepackage{graphicx}
\usepackage{textcomp}
\usepackage{xcolor}
\usepackage{booktabs, multirow,array}
\usepackage{float}
\usepackage{subfig}

\def\BibTeX{{\rm B\kern-.05em{\sc i\kern-.025em b}\kern-.08em
    T\kern-.1667em\lower.7ex\hbox{E}\kern-.125emX}}
\begin{document}

\title{Beyond Conic Section Mainlobe Coverage for Unmanned Aerial Vehicle
}
\makeatletter
\newcommand{\linebreakand}{%
  \end{@IEEEauthorhalign}
  \hfill\mbox{}\par
  \mbox{}\hfill\begin{@IEEEauthorhalign}
}
\makeatother
\author{\IEEEauthorblockN{Yi Geng}
	\IEEEauthorblockA{\textit{Unisoc Research} \\
		Nanjing, China \\
		yi.geng@unisoc.com}
	\and
	\IEEEauthorblockN{Sebastian Euler}
	\IEEEauthorblockA{\textit{Ericsson Research} \\
		Stockholm, Sweden \\
		sebastian.euler@ericsson.com}
}
\maketitle

\begin{abstract}
The cellular-connected drone market is one of the most promising markets of 5G. However, it still accounts for a small share of the overall telecommunication market, and its share is unlikely to increase significantly in the foreseeable future. Deploying dedicated networks with up-tilted antennas can be an option, but the monetary cost of dedicated networks directly impacts the acceptance of mobile operators. Therefore, cost-efficient aerial coverage solutions must be developed. Reusing a network for terrestrial coverage is a cost-efficient approach for aerial coverage, but several critical challenges caused by antenna sidelobes should be solved. In this paper, a novel method for aerial coverage is proposed. By tweaking the measurement report handling mechanism, signals from sidelobes reported by drones above a predefined height can be identified and ignored. Simulation results show that the conventional cellular network with the proposed method can provide wide and continuous aerial coverage with satisfactory quality.
 
\end{abstract}


\section{Introduction}
The connected sky is developing into an indispensable part of the Internet of Things (IoT): anywhere, anytime, anything. With this trend, cellular networks are expected to provide wide area, high quality, ubiquitous coverage, and secure connectivity for unmanned aerial vehicles (UAVs, aka drones). Deployment of dedicated networks for aerial coverage with up-tilted antennas and separate spectrum is essentially the ideal option and can provide satisfactory aerial coverage due to free of interference from terrestrial user equipments (UEs) and dedicated radio resources. However, designing, deploying, and optimizing another network with adequate coverage in the airspace are not cost-efficient solutions due to the fact that most of the traffic will be on the terrestrial level even in the near future. According to the prediction made by GSMA, the total number of UAVs will reach 86.5~million by 2025 \cite{b1}. Meanwhile, terrestrial users and IoT connections will reach 5.8~billion and 25~billion by 2025, respectively \cite{b2}. Therefore, dedicated networks for aerial coverage have not been widely deployed in practice so far. Reusing the networks for terrestrial coverage, that is, deploying aerial coverage in the already existing cellular networks has been identified as the most suitable way to implement early aerial coverage and facilitate the rapid growth of the UAV ecosystem \cite{b3}.

\section{Characteristics and Challenges of Aerial Coverage}

The ground-to-air radio channels between base station (BS) antennas and UAVs can be classified into line-of-sight (LOS) channels and non-LOS channels. The channel models of UAVs in the sky have shown different propagation characteristics compared to the radio channels of terrestrial UEs \cite{b4}. As height increases, the probability of LOS channels between UAVs and BSs significantly increases since obstacles between UAVs and BS antennas occur less frequently. Reflection- and diffraction-induced propagation losses, which highly attenuate the signals in NLOS environments, do not exist in LOS environments. As a result, the high signal level of a LOS channel over a long distance in the sky is predictable. This fact has been verified by field measurements, showing that a 5G macro BS can provide aerial coverage up to 6-8~km by LOS path at the height of 300~m \cite{b5}.

Another unexpected effect making the aerial coverage different from the terrestrial coverage is caused by antenna sidelobes. Antenna arrays emit unwanted radiation in undesired directions, known as sidelobes. The BS antennas in conventional cellular networks are designed and optimized for serving terrestrial users. With an appropriate tilt angle, antenna mainlobes are pointing downwards to the ground. Within such cellular networks, UAVs flying above the antenna height may be covered by a large number of narrow sidelobes pointing upwards. As a result, connected UAVs in the sky may suffer from two major challenges:

\subsection{Scattered cell association pattern}

\begin{figure}[t]
	\centering
	\subfloat[0 m]{\label{fig:a}\includegraphics[width=0.48\columnwidth]{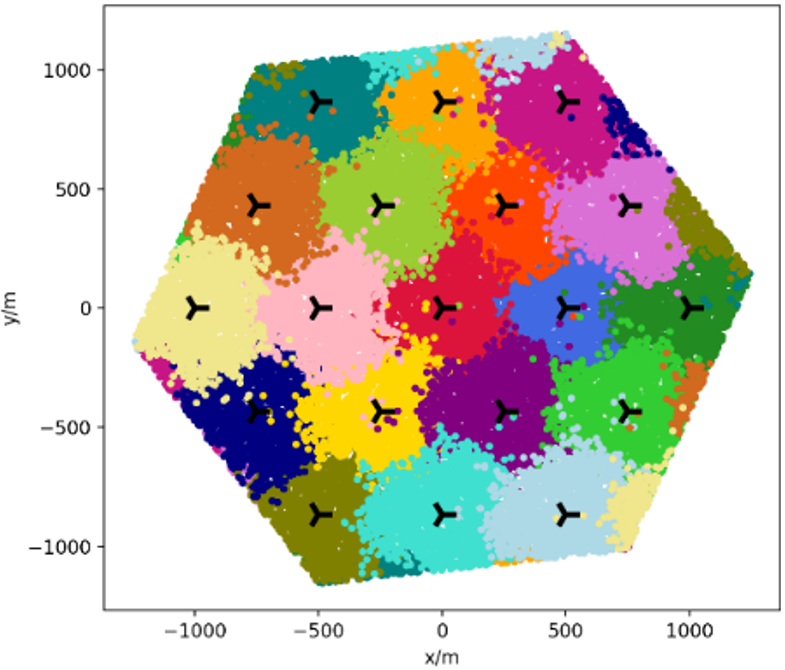}}\quad
	\subfloat[300 m]{\label{fig:b}\includegraphics[width=0.48\columnwidth]{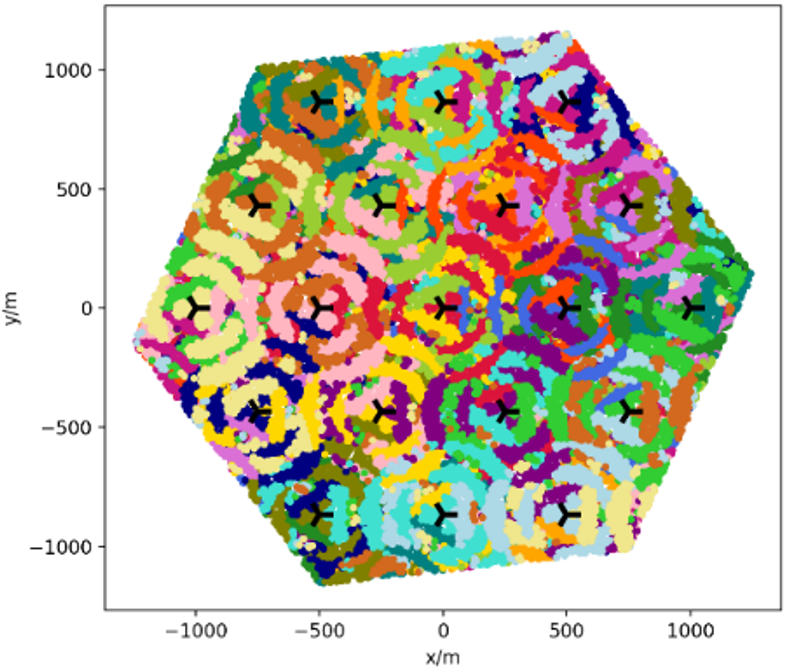}}\\	
	\caption{The cell association patterns at the height of (a) 0~m and (b) 300~m, the color at a certain position represents the cell that can provide the maximum RSS.}
\end{figure}

A large number of narrow sidelobes emitting into the airspace result in the phenomenon of scattered cell association patterns, which are particularly noticeable in the airspace above the BS antenna height. Fig.~1 shows the simulation results of cell association patterns based on the maximum received signal strength (RSS) at the ground level and the height of 300~m in a hexagonal network with 19 three-sector BSs \cite{b5}. The color at a certain position in Fig.~1 represents the BS that can provide the maximum RSS for that position. At ground level, the antenna mainlobes provide large and continuous coverage areas. These mainlobe coverage areas shape the cell association pattern as illustrated in Fig.~1(a). The best cell is most often the one closest to the terrestrial UEs, hence this is a suitable pattern for terrestrial coverage. However, at the height of 300~m, the multiple narrow antenna sidelobes create scattered cell association patterns, and the best cell of the UAV at 300~m may no longer be the closest one, as illustrated in Fig.~1(b).

\subsection{Poor aerial mobility performance}

The simulations and field trial results in \cite{b4} indicate that the mobility performance of UAVs in the airspace is significantly worse than that of terrestrial users at ground level, including handover rate, handover failure (HOF) rate, and radiolink failure (RLF) rate. For example, the HOF rate of connected UAVs with a velocity of 160~km/h at 300~m is as high as 97\% \cite{b4}. The discrepancy in mobility performance between ground level and height of 300~m is mainly due to the different radiation patterns of mainlobes and sidelobes. The mainlobe is typically emitted horizontally or downwards with a few degrees, the mainlobe thus provides a wide coverage area for terrestrial users. When a terrestrial user is leaving the mainlobe coverage area of its serving cell, the received signal from the serving cell gradually degrades due to the path loss over increasing distance. The terrestrial UE can make handover to a better neighbor cell when the signal strength of the serving cell gets worse than a predefined threshold. Unlike the mainlobe emitting in the horizontal direction, the sidelobe is emitted upwards to the sky, therefore, the sidelobe direction is almost perpendicular to the direction of the motion of UAVs flying horizontally. When a connected UAV with high horizontal velocity leaves the field of view (FOV) of a sidelobe, the UAV will immediately experience very high RSS degradation. In other words, the sidelobe has a sharp coverage edge in the horizontal direction. Fig.~2 shows simulated RSRP measured by a UAV moving away from the sidelobe coverage of the UAV's serving cell at 300~m. At 0~s, the UAV is served by cell~0. After 1~s, RSRP of the serving cell begins to drop rapidly when the UAV leaves the sidelobe coverage of its serving cell. Due to the sharp fading of the signal from the serving cell, the default mobility procedures might be too slow for successful handover execution to any neighbor cells (cell~1 to cell~6), leading to HOF at 3.5~s.

\begin{figure}[t]
	\centerline{\includegraphics[width=\linewidth, height=10cm, keepaspectratio]{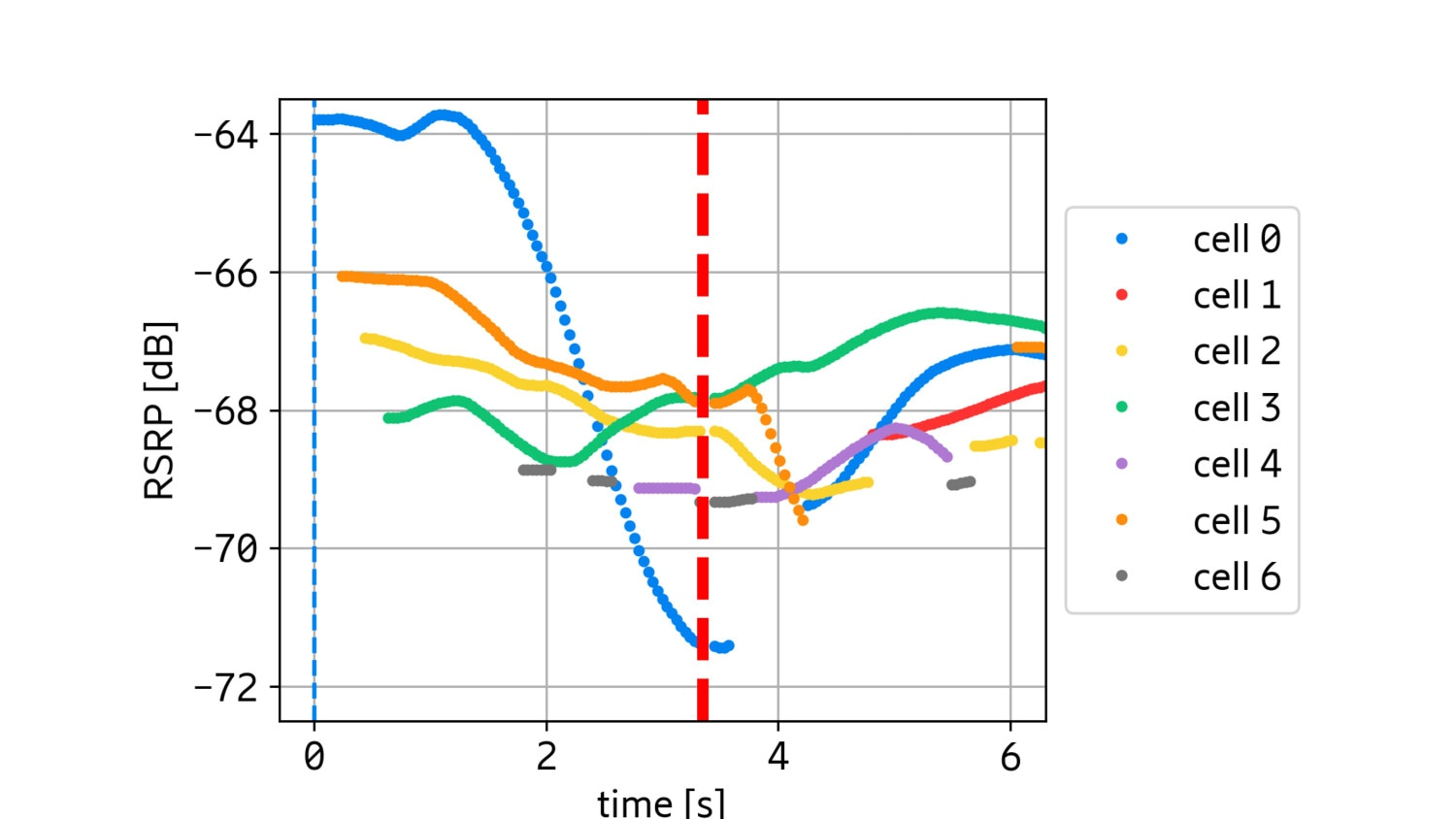}}
	\caption{Sudden drop in signal strength when a UAV at 300~m leaves the sidelobe coverage.}
	\label{fig_2}
\end{figure}

\section{Beyond Conic Section Mainlobe Coverage in Airspace}

\subsection{Method}

As mentioned in Section~II, existing cellular networks for terrestrial coverage purposes cannot provide satisfactory aerial coverage, especially at 300~m and above. The challenges of aerial coverage resulting from sidelobes can be solved by ensuring that the signals from sidelobes are ignored by the measurement evaluation and handling algorithm of the networks. If this can be accomplished, the UAV will not make handover to sidelobes, even if the signals from sidelobes reported by the UAV are better than the signal from the serving cell. Accordingly, handover will not be triggered frequently in the airspace, compared to the case where UAVs are served by scattered sidelobes. Even the worse mobility performance due to sharp fading in signal strength is solved since mainlobes emitting horizontally do not have a sharp coverage edge in the vertical direction.

In this paper, we define a new concept ``beyond-conic-section (BCS) mainlobe coverage''. Fig.~3 illustrates a directional antenna with a tilt angle of 0\textdegree. H denotes the antenna height, and the apex angle of the mainlobe cone is 30\textdegree. The conic section is the curve obtained as the intersection of the mainlobe cone with the horizontal plane at 300~m. UAVs at 300~m beyond the conic section are within the FOV of the mainlobe. In other words, UAVs in the area colored in green can be covered by the mainlobe, we call this area ``BCS mainlobe coverage''. According to trigonometry, the distance from the apex of the conic section at 300~m to the vertical axis of the antenna is (300-H)/tan15\textdegree, which is the minimum range of BCS mainlobe coverage. 

\begin{figure}[t]
	\centerline{\includegraphics[width=\linewidth, height=10cm, keepaspectratio]{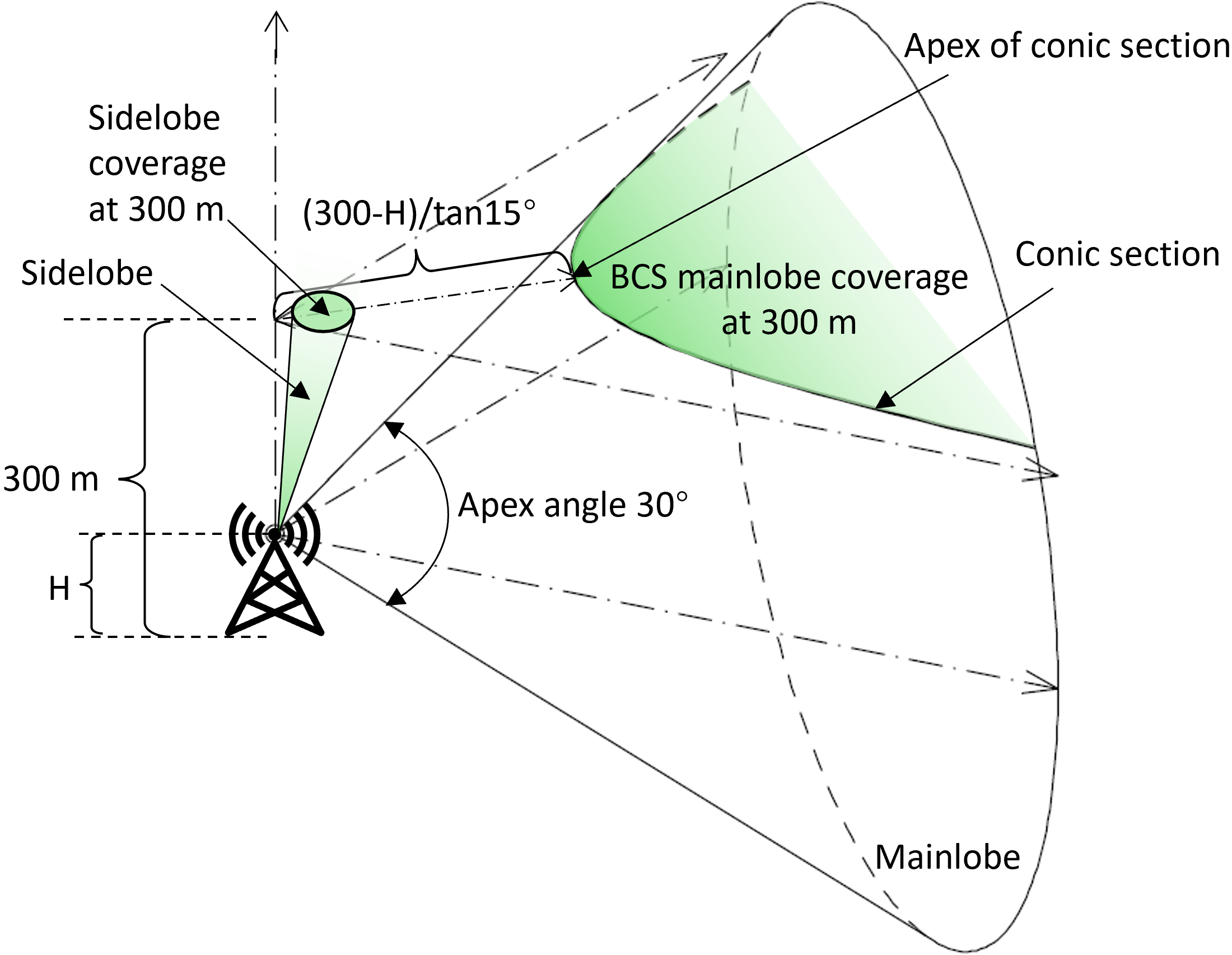}}
	\caption{An illustration of the geometry of aerial coverage provided by the mainlobe of a BS antenna.}
	\label{fig_3}
\end{figure}

\begin{figure}[t]
	\centerline{\includegraphics[width=\linewidth, height=10cm, keepaspectratio]{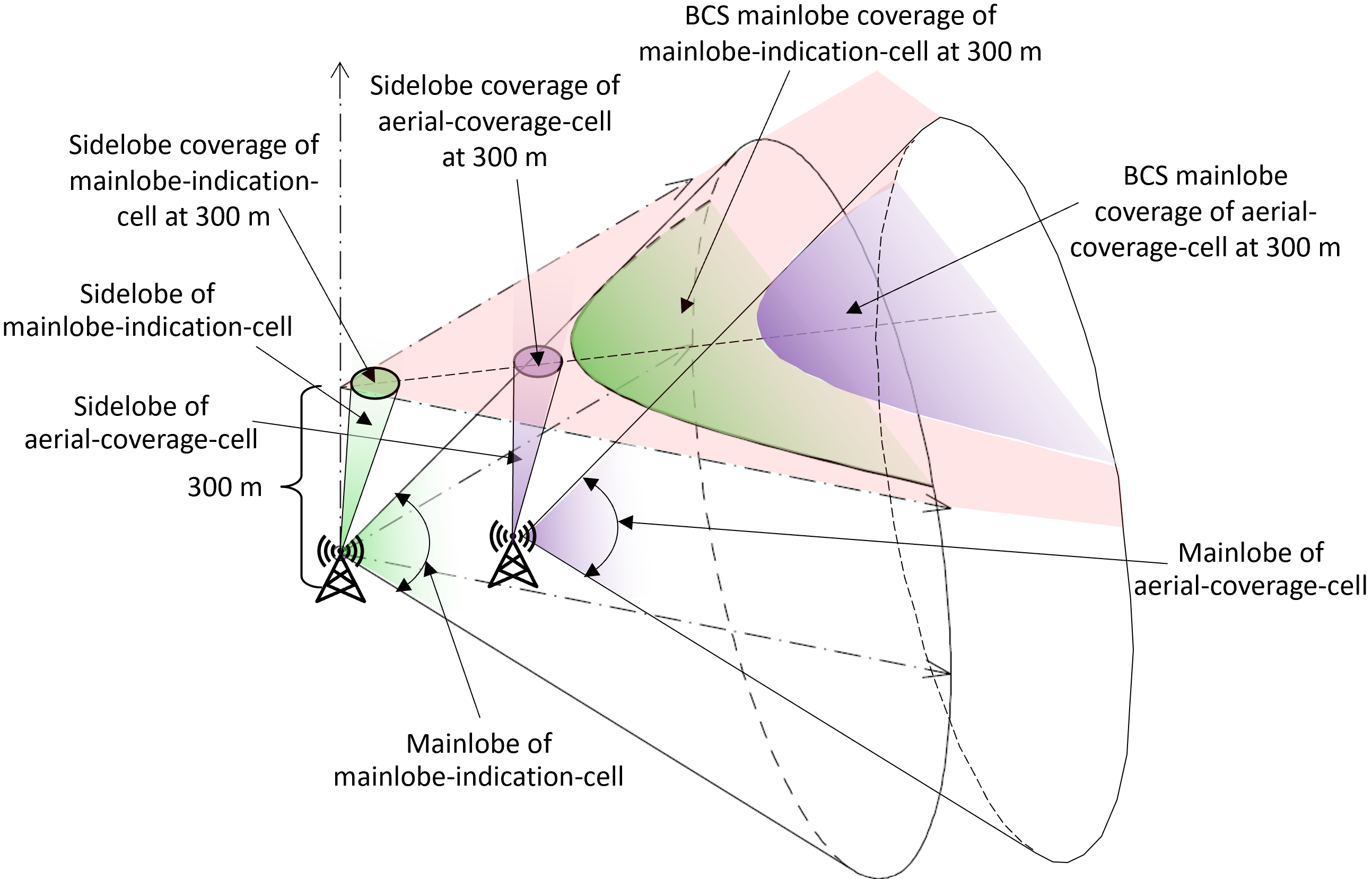}}
	\caption{Almost overlapping BCS mainlobe coverage areas and non-overlapping sidelobe coverage areas provided by an aerial-coverage-cell and a mainlobe-indication-cell in a cell-pair.}
	\label{fig_4}
\end{figure}

\begin{figure}[t]
	\centerline{\includegraphics[width=\linewidth, height=10cm, keepaspectratio]{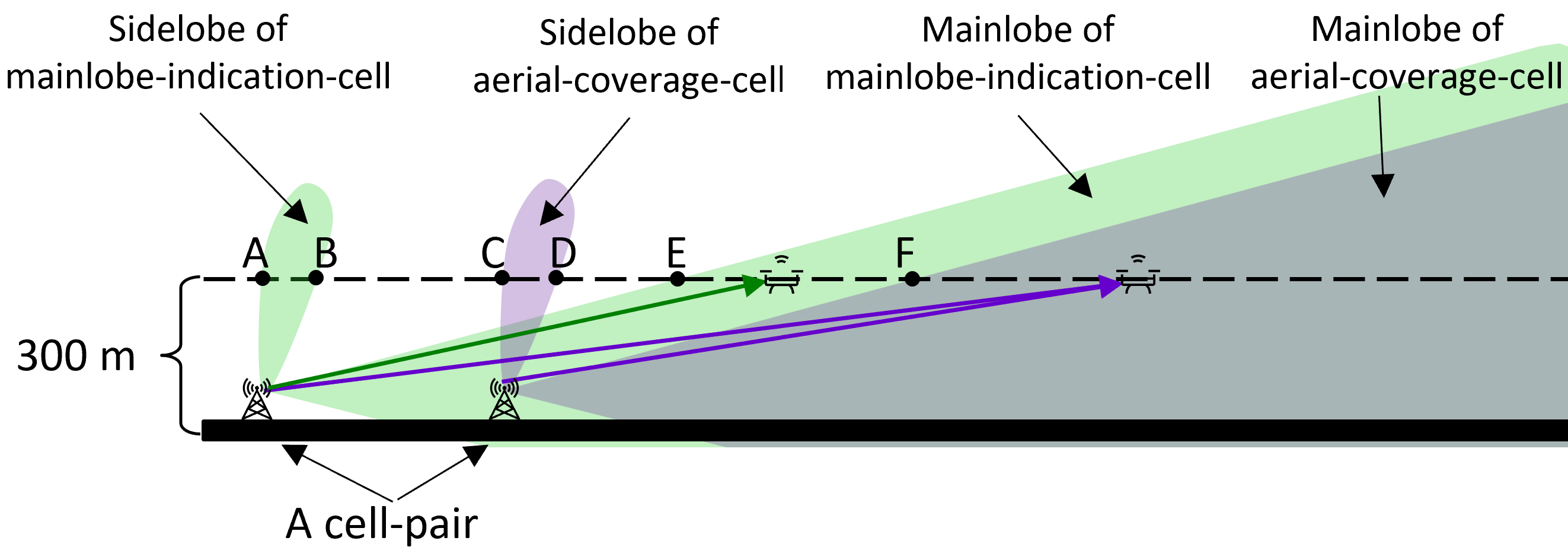}}
	\caption{Geometrical illustration of the lobes associated with a pair of mainlobe-indication-cell and aerial-coverage-cell.}
	\label{fig_5}
\end{figure}

Based on BCS mainlobe coverage, we propose a novel method to circumvent the impact from sidelobes for aerial coverage as follows \cite{b6}. In a cellular network for covering airspace and ground simultaneously, a set of cells are selected to provide aerial coverage above a predefined height (e.g., 300~m), named aerial-coverage-cell. Another set of cells are selected for indicating the mainlobe steering range of the aerial-coverage-cells, named mainlobe-indication-cell, while other cells are bypassed by the proposed method. It is worth noting that aerial-coverage-cells and mainlobe-indication-cells can provide terrestrial coverage also, they just need to be identified by the measurement evaluation and handling algorithm in the proposed method. An aerial-coverage-cell in purple and a mainlobe-indication-cell in green are combined into a cell-pair as illustrated in Fig.~4. The aerial-coverage-cell and the mainlobe-indication-cell in a cell-pair should be selected from two adjacent BSs respectively and both cells should have similar antenna boresight (mainlobe direction). As a result, the aerial-coverage-cell and the mainlobe-indication-cell in a cell-pair can provide almost overlapping BCS mainlobe coverage areas and non-overlapping sidelobe coverage areas at 300~m as illustrated in Fig.~4. Fig.~5 illustrates the geometrical side view of the lobe patterns associated with a pair of aerial-coverage-cell and mainlobe-indication-cell. The Boolean intersection of the aerial coverage provided by the aerial-coverage-cell and the mainlobe-indication-cell at 300~m is the coverage area where the UAVs can measure and report signals from both cells in the cell-pair. The Boolean intersection of the aerial coverage provided by both cells at 300~m is the area beyond point F as illustrated in Fig.~5. That means a UAV at 300~m on the right-hand side of \mbox{point F} can measure and report the signals from both the aerial-coverage-cell and the mainlobe-indication-cell in the cell-pair. On the left-hand side of point F, the UAV at 300~m can measure and report signals from one cell in the cell-pair at most. For example, within AB, only the signals from the sidelobe of the mainlobe-indication-cell can be measured; within BC and DE, none of the cells can be measured; within EF, only the signals from the mainlobe of the mainlobe-indication-cell can be measured.

The proposed method follows a stepwise procedure as illustrated in Fig.~6 and is described as follows. After having received a measurement report (MR) that is triggered by a handover event (e.g., event A3), the network with the proposed method evaluates and handles the MR according to the below conditions: if the MR is reported by a UE above the predefined height, and the best cell in the MR is an aerial-coverage-cell, and its pair mainlobe-indication-cell in the cell-pair is included in the MR also, then a handover decision is made to handover the UE from the serving cell to the best cell in the MR. Otherwise, the MR is ignored even if the signal level of the best cell in the MR is stronger than the signal level of the serving cell.

\begin{figure}[t]
	\centerline{\includegraphics[width=\linewidth, height=10cm, keepaspectratio]{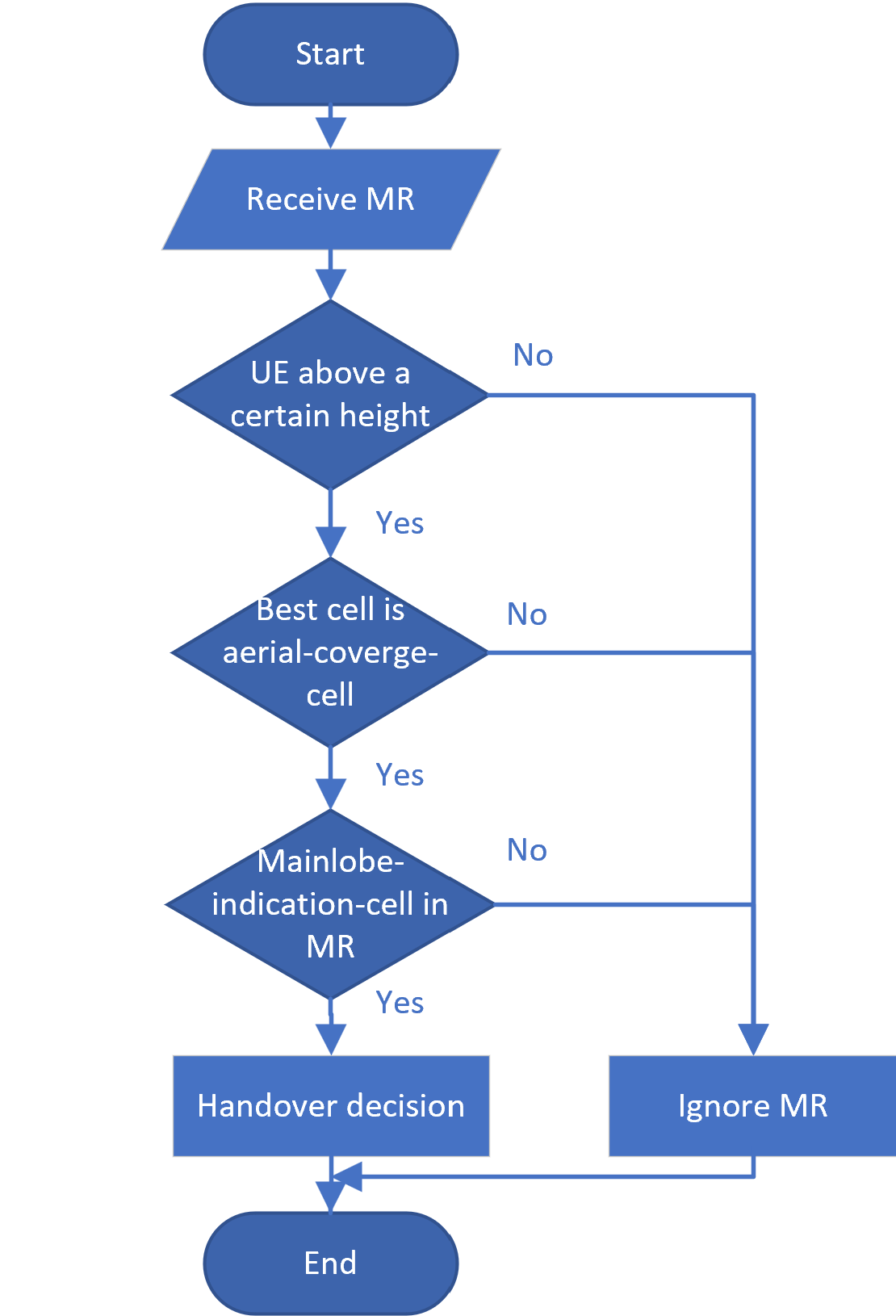}}
	\caption{Stepwise procedure of the proposed method.}
	\label{fig_6}
\end{figure}

Based on the aforementioned method, Fig.~7 gives an example of how the network makes handover decisions for a UAV flying from left to right along the dashed line at 300~m. In this network, cell-1 and cell-2 are combined into \mbox{cell-pair-1}, cell-4 and \mbox{cell-5} are combined into cell-pair-2. \mbox{Cell-2} and \mbox{cell-5} in purple are selected as aerial-coverage-cells, cell-1 and cell-4 in green are selected as mainlobe-indication-cells, and \mbox{cell-3} in yellow is a conventional cell in the network. We assume that the UAV has connected to cell-2 at position G already. At position H, the signal from the sidelobe of cell-3 is stronger than the signal from the serving cell (signal from the mainlobe of cell-2), hence the UAV reports MR event A3 at position H. Similarly, the UAV reports MR event A3 at positions I, J, and K, respectively. Because cell-3 and cell-4 are not aerial-coverage-cells, the MR reported at position H and I are ignored. This is true also for the MR at position J, even though \mbox{cell-5} is an aerial-coverage-cell, but its pair mainlobe-indication-cell (cell-4) is not included in the MR since position~J is beyond the FOV of both mainlobe and sidelobe of cell-4. Therefore, MRs at H, I, and J cannot trigger a handover decision. At position~K, the UAV reports cell-5 is stronger than the serving cell and its pair mainlobe-indication-cell (cell-4) is included in the same MR also, so that the network makes a decision to handover the UAV from cell-2 to cell-5. This example indicates that by the proposed method, only the signals from mainlobes of aerial-coverage-cells can be the candidates for serving UAVs above the predefined height. Signals from sidelobes, no matter which cells emit the sidelobes, e.g., sidelobes of aerial-coverage-cells, mainlobe-indication-cells, and other conventional cells, are ignored by the network.

It is worth noting that 3GPP Release~15 has introduced MR events H1 and H2 for connected UAVs \cite{b7}, which are triggered when the height of the UAV is above or below a configured height threshold. With the aid of events H1 and H2, the proposed method can be applied to UAVs above a predefined height only and does not impact terrestrial UEs. Moreover, the cell association patterns in the sky depend on many parameters, such as inter-site-distance, antenna type, antenna tilt, antenna height, etc. For example, if we change antenna tilt angle or select another cell as mainlobe-indication-cell, the aerial coverage pattern will change to a totally different one. Therefore, we could flexibly select the mainlobe-indication-cell and aerial-coverage-cell in the network to meet various aerial coverage requirements.

\begin{figure*}[t]
	\centerline{\includegraphics[width=\linewidth, height=10cm, keepaspectratio]{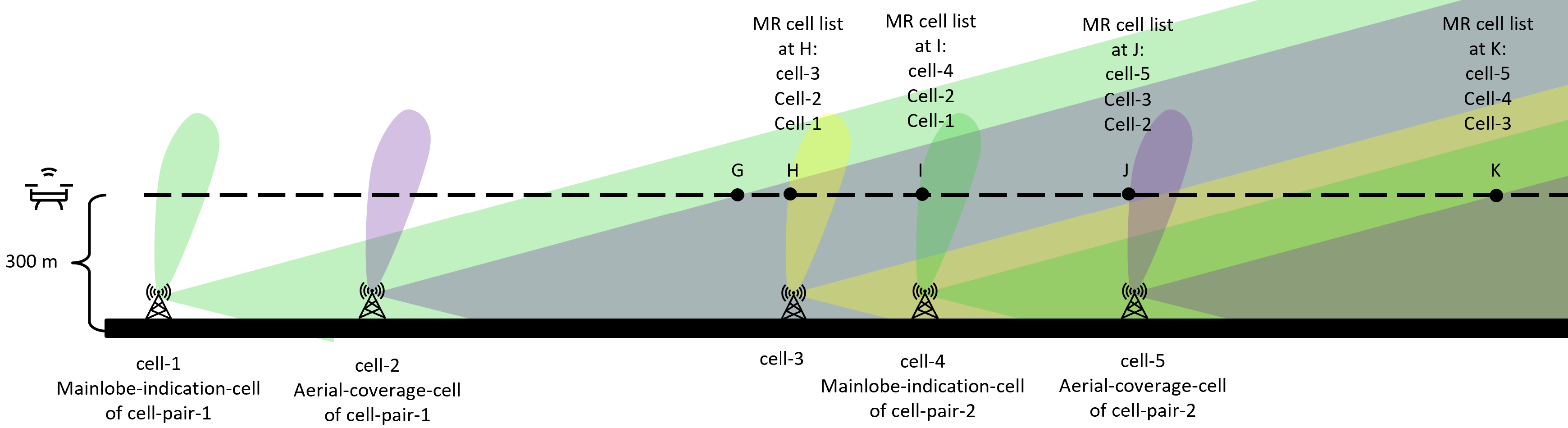}}
	\caption{An illustration of handover decisions made for a UAV flying along a flight path at 300~m.}
	\label{fig_7}
\end{figure*}

\subsection{Simulations}
In order to validate the proposed method, simulations of aerial coverage at 300~m provided by a homogeneous hexagonal-grid network with the proposed method have been carried out. The network consists of 19 three-sector BSs with the antenna tilt angle of 0\textdegree. The inter-site-distance and antenna height are set to 500~m and 75~m, respectively. The half-power beam width (HPBW) of the mainlobe in the horizontal and vertical planes are 120\textdegree~and 30\textdegree, respectively.

In principle, the mainlobe-indication-cell should be selected around the aerial-coverage-cell. Meanwhile, the mainlobe direction of the mainlobe-indication-cell should coincide with the mainlobe direction of the aerial-coverage-cell. Then the two cells in a cell-pair can provide almost overlapping mainlobe coverage areas and non-overlapping sidelobe coverage areas in the airspace, as the proposed method requires. For example, in Fig.~8, we select \mbox{cell-6} as the aerial-coverage-cell. Under these circumstances, \mbox{cell-7}, \mbox{cell-8}, and \mbox{cell-9} can be candidates for the mainlobe-indication-cell of \mbox{cell-6}. If we select \mbox{cell-7} as the mainlobe-indication-cell, the path gain pattern in Fig.~8 indicates the path gain of \mbox{cell-6} and \mbox{cell-7} at 300~m. For clarity of illustration, other cells are inactive. It is evident that at 300~m \mbox{cell-6} and \mbox{cell-7} provide almost overlapping mainlobe coverage areas inside the red dashed circle, and the path gain of mainlobe coverage at 300~m is about -100~dB. Furthermore, the two cells provide non-overlapping sidelobe coverage areas, which are the two separate areas in yellow and green. The signal level of the sidelobe coverage is in excess of 10~dB than that of the mainlobe coverage. This is because signals emitted from mainlobes propagate over a long distance to the height of 300~m, and the longer distance leads to relatively higher path loss.

As shown in Fig.~9, we select cell-A, cell-C, and \mbox{cell-E} as aerial-coverage-cells in the hexagonal network with 19 three-sector BSs. Cell-B, cell-D, and cell-F are selected as the mainlobe-indication-cells of cell-A, cell-C, and cell-E, respectively. We can see that by implementing the proposed method, the BCS mainlobe coverage areas of cell-A, cell-C, and cell-E, which are colored in dark yellow, cyan, and green, respectively, form a wide and continuous cell association pattern at 300~m, while the strong and scattered sidelobe signals are ignored.

The proposed method ensures that the UAVs above the predefined height are connecting to the mainlobes of BS antennas constantly, even though the sidelobes may provide stronger signals for UAVs. This method avoids the mobility problems caused by the scattered cell association patterns and sudden signal strength drops in the airspace. Thus, the proposed method reduces the handover rate and improves the HOF rate of UAVs. In addition, it is a cost-efficient solution for aerial coverage without any hardware changes or investments, but by minor software changes to enhance the network that is anyway deployed for terrestrial coverage. However, the proposed method requires that the BS antennas are not tilted downwards too much. If the down-tilt angle is too large, the BSs cannot provide BCS mainlobe coverage at larger heights and the method isn't effective any more. This problem can be mitigated by using antenna configurations with large vertical HPBW, e.g., antenna configurations for covering high-rise buildings in dense urban deployments. For such antenna configurations, the proposed method is more robust, and hence still feasible even when the BS antennas are tilted downward for terrestrial coverage.

\begin{figure}[t]
	\centerline{\includegraphics[width=\linewidth, height=10cm, keepaspectratio]{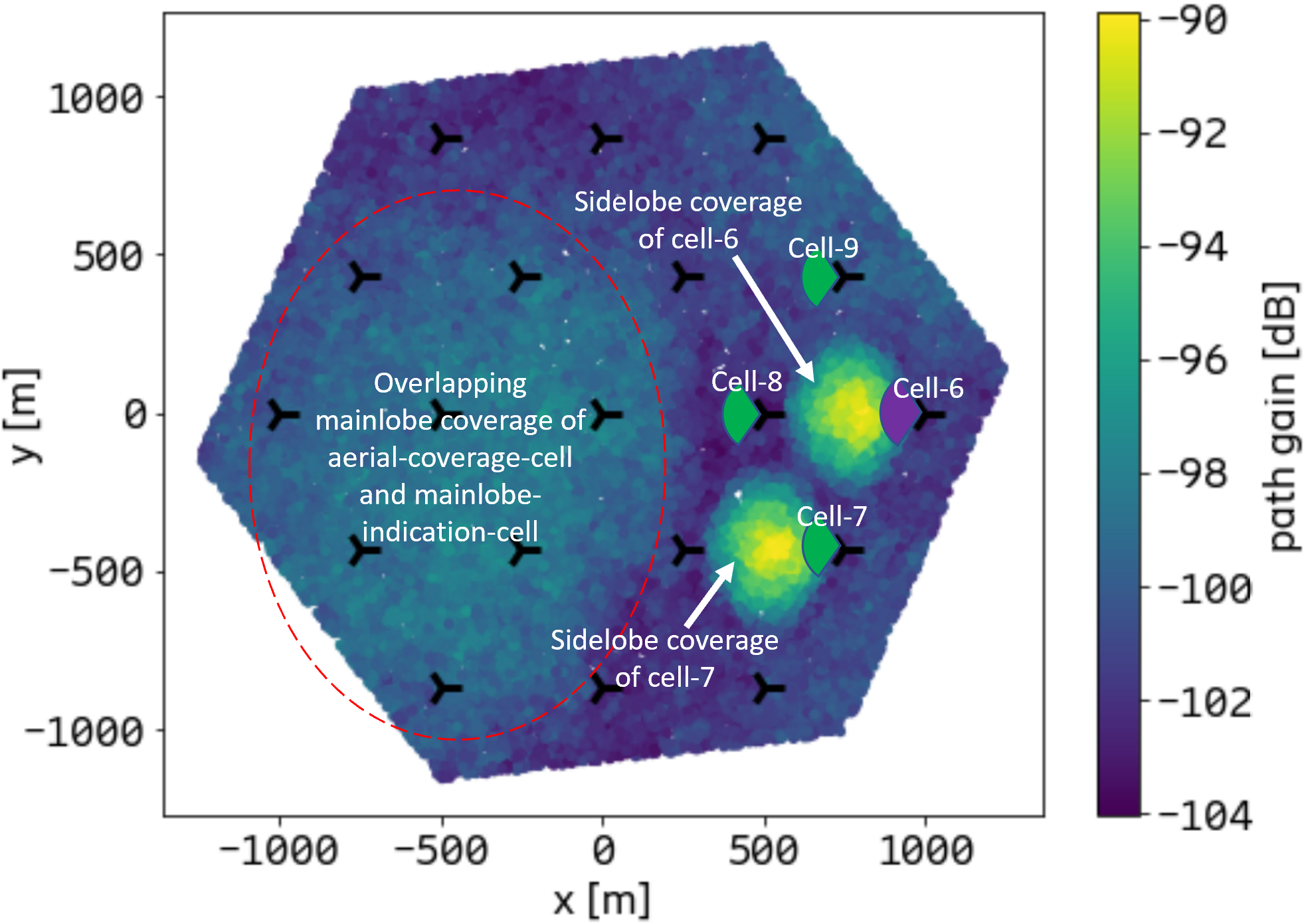}}
	\caption{Mainlobe coverage and sidelobe coverage provided by two cells in a cell-pair and path gain between BS antennas and UAVs at 300~m.}
	\label{fig_8}
\end{figure}

\section{Aerial-Coverage-Cell Group}
Compared to the scattered aerial coverage provided by the conventional cellular network as illustrated in Fig.~1(b), the method proposed in Section~III can overcome sidelobe-related challenges including scattered cell association patterns and poor mobility performance. However, the method has two significant drawbacks. First, the capacity of aerial coverage is limited as only a small fraction of cells are selected as aerial-coverage-cells. As shown in Fig.~9, three cells out of a total of 57~cells are selected as aerial-coverage-cells in the network, hence only about 5\% of the total network capacity can be used for aerial coverage. Second, the proposed method cannot provide flexible QoS for various UAV use cases. Cellular networks are expected to provide flexible and differentiated QoS for UAVs \cite{b8}. For instance, the UAV command and control (C2) link does not require a very high data rate, but low latency, ubiquitous coverage, extremely high reliability, and seamless mobility for traffic safety, e.g., preventing UAVs from falling to the ground or avoiding mid-air collisions. Unlike C2 link, UAV applications for high-definition video broadcast and surveillance require UAVs to transmit tremendous amounts of uplink (UL) data and show an obvious asymmetry between UL and downlink (DL) data transmission.

To overcome these drawbacks, a new notion ``aerial-coverage-cell group'' is proposed. The basic idea is to divide the selected aerial-coverage-cells into groups according to the QoS requirements of various UAV applications. As shown in Fig.~9, an aerial-coverage-cell group comprising aerial-coverage-cells cell-A, C, and E (abbreviated to \mbox{group-1} below) provides BCS mainlobe coverage at 300~m. In the same network, we can create another aerial-coverage-cell group comprising aerial-coverage-cells cell-B, D, F, G, H, and I (abbreviated to \mbox{group-2} below), which produces a totally different BCS mainlobe coverage pattern at 300~m as illustrated in Fig.~10. It is worth noting that the proposed method allows a cell to play different roles in different groups. For instance, for group-1, cell-B is combined with cell-A into a cell-pair, and cell-B is the mainlobe-indication-cell of cell-A. However, in group-2, \mbox{cell-B} acts as an aerial-coverage-cell to provide aerial coverage.

The aerial-coverage-cell group is a network structure that enables multiple groups for aerial coverage in the same network infrastructure. From the perspective of UAV applications, these groups isolate the network as if the UAVs are served by independent radio networks, that is, a UAV above a predefined height can perform handovers between the aerial-coverage-cells belonging to the same group only. In order to fulfill the diverse requirements requested by different UAV applications, each group can be configured and optimized in a cell-specific manner for UAV applications with different QoS requirements. For example, as shown in Fig.~9 and Fig.~10, the coverage size of each aerial-coverage-cell in group-1 is two times larger than that in group-2, thus inducing a low handover rate. High capacity for UAVs is however unlikely to be available due to the fact that there are only three aerial-coverage-cells in group-1. By cell-level parameter settings and configurations in \mbox{group-1} for low-latency and high-reliability purposes, the connection robustness of UAV can be improved, hence \mbox{group-1} can be more suitable for supporting UAV applications that require uninterrupted and robust data exchange but low data rate, e.g., C2 communication for operating UAVs. On the other hand, thanks to the high density of aerial-coverage-cells in group-2, group-2 can provide 100\% increase in capacity compared to group-1 for the same coverage area. Therefore, group-2 can be used for supporting UAV-related mobile broadband applications (e.g., high-definition video broadcast and surveillance). This can be achieved by allocating most of the time and frequency resources in the UL direction (high UL/DL ratio) for the aerial-coverage-cells in group-2.

\begin{figure}[t]
	\centerline{\includegraphics[width=\linewidth, height=10cm, keepaspectratio]{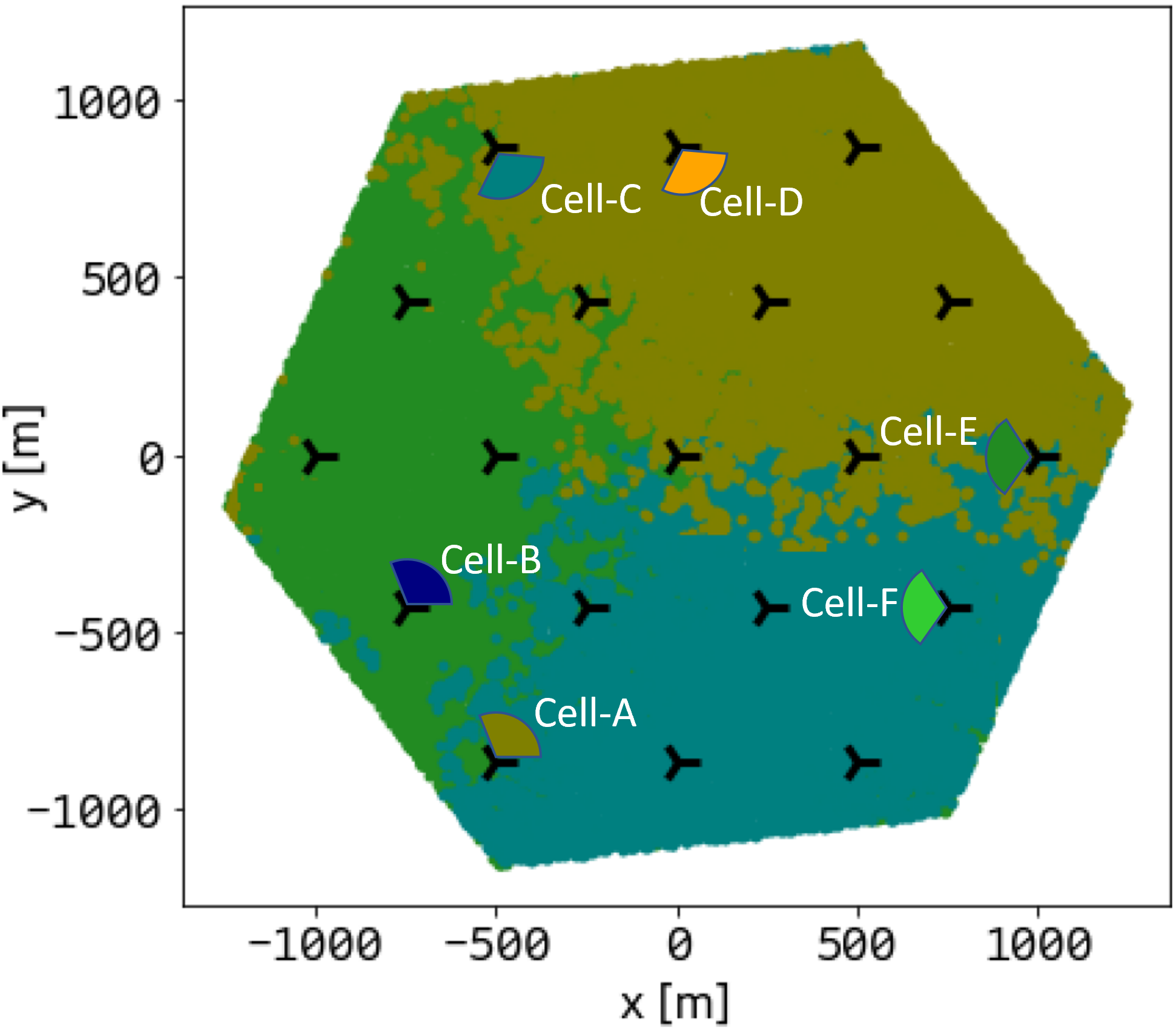}}
	\caption{Cell association pattern at 300~m provided by three aerial-coverage-cells (cell-A, C, and E) in a hexagonal network with 19 three-sector BSs.}
	\label{fig_9}
\end{figure}

\begin{figure}[t]
	\centerline{\includegraphics[width=\linewidth, height=10cm, keepaspectratio]{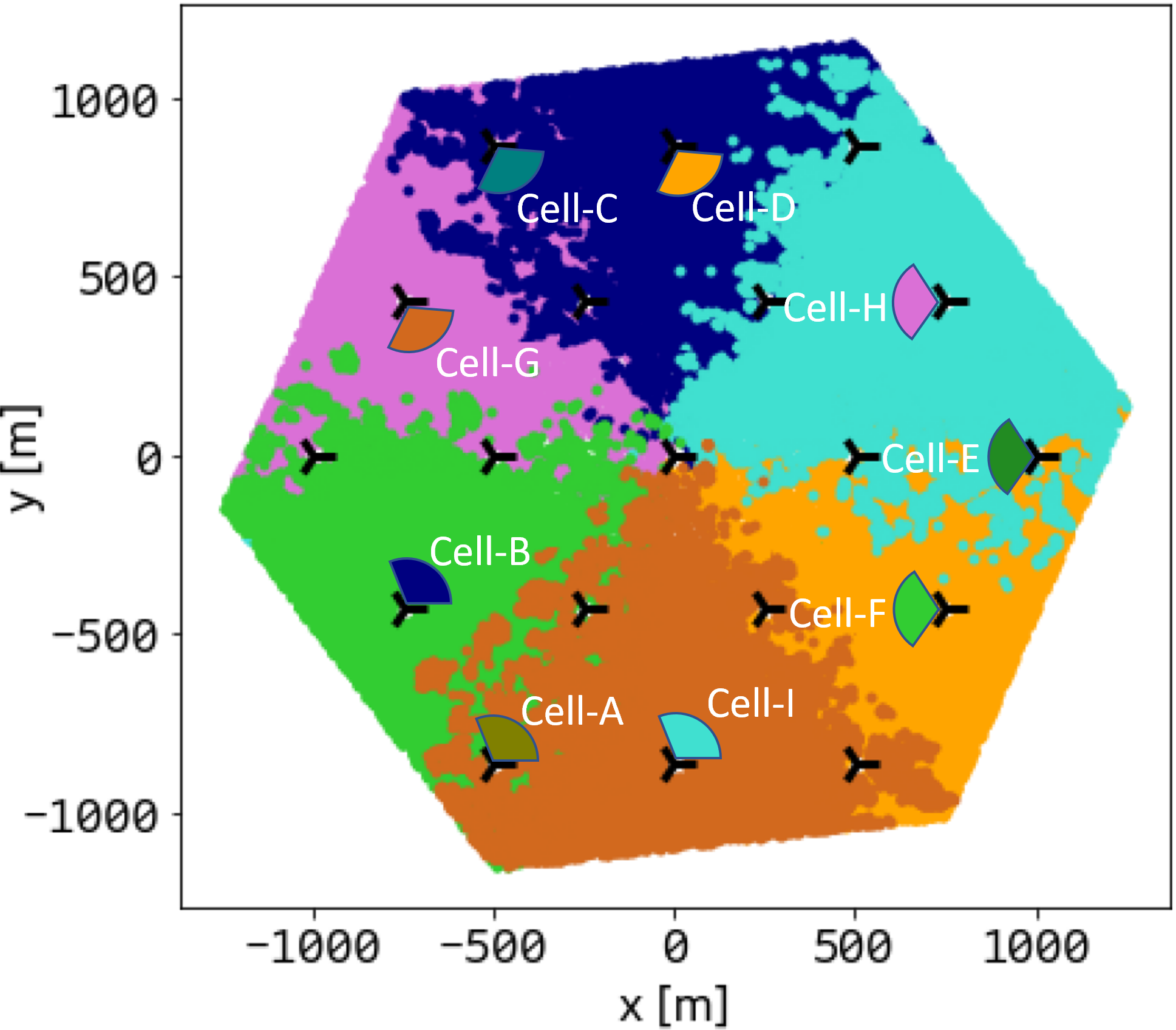}}
	\caption{Another cell association pattern at 300~m provided by six aerial-coverage-cells (cell-B, D, F, G, H, and I) in the same network depicted in Fig.~9.}
	\label{fig_10}
\end{figure}

\section{Conclusions}
In this paper, we provide a geometrical model describing the spatial relationship between the mainlobes and the sidelobes of two cells in a cell-pair. Based on this geometrical model, an aerial coverage method is proposed. The method uses a stepwise measurement evaluation and handling algorithm to identify and ignore signals from sidelobes reported by UAVs above a predefined height. Simulation results show that the proposed method can provide wide and continuous aerial coverage patterns and avoid the scattered cell association patterns caused by antenna sidelobes. Moreover, by creating multiple aerial-coverage-cell groups, varying QoS requirements for UAV applications can be supported.

\vspace{12pt}
\color{red}
\end{document}